\documentclass[12pt,draftclsnofoot,journal,onecolumn]{IEEEtran}
\usepackage{amsmath,graphicx}
\usepackage{times}
\usepackage[utf8]{inputenc}
\usepackage{cite}
\usepackage{floatrow}
\renewcommand*{\thesubsubsection}{\Alph{subsubsection}}
\usepackage{amsmath,amssymb,amsfonts,stfloats}
\usepackage{tabularx}
\usepackage{graphicx}
\usepackage{textcomp}
\usepackage{floatrow}
\usepackage{xcolor}
\usepackage[hyphens]{url}
\usepackage[font=small]{caption}
\usepackage{subcaption}
\usepackage{booktabs}

\usepackage{multicol}
\usepackage{multirow}
\usepackage{acronym}
\usepackage{url}
\usepackage{mathtools}
\usepackage{algorithm}
\usepackage{algpseudocode} 
\usepackage{enumitem}
\usepackage[hang,flushmargin]{footmisc}
\usepackage[caption=false,font=footnotesize]{subfig}

\def\BibTeX{{\rm B\kern-.05em{\sc i\kern-.025em b}\kern-.08em
    T\kern-.1667em\lower.7ex\hbox{E}\kern-.125emX}}

\title{\huge{A Multi-Modal Simulation Framework to Enable\\Digital Twin-based V2X Communications\\in Dynamic Environments}}

\author{
\IEEEauthorblockN{Lorenzo~Cazzella, Francesco~Linsalata, Maurizio~Magarini,\\Matteo~Matteucci, and~Umberto~Spagnolini}

}

\begin{document}
\maketitle
\begin{abstract}

Digital Twins (DTs) for physical wireless environments have been recently proposed as accurate virtual representations of the propagation environment that can enable multi-layer decisions at the physical communication equipment. At high-frequency bands, DTs can help to overcome the challenges emerging in high mobility conditions featuring vehicular environments. In this paper, we propose a novel data-driven workflow for the creation of the DT of a Vehicle-to-Everything (V2X) communication scenario and a multi-modal simulation framework for the generation of realistic sensor data and accurate mmWave/sub-THz wireless channels. The proposed method leverages an automotive simulation and testing framework and an accurate ray-tracing channel simulator.
Simulations over an urban scenario show the achievable realistic sensor and channel modelling both at the infrastructure and at ego-vehicles. We showcase the proposed framework on the DT-aided blockage handover task for V2X link restoration, leveraging the framework's dynamic channel generation capabilities for realistic vehicular blockage simulation.

\end{abstract}


\section{Introduction}\label{sect:introduction}

The sixth generation (6G) technology targets to revolutionize the mobility industry enhancing the role of wireless connections in car manufacturing. Indeed, connected vehicles are based on the vehicle-to-everything (V2X) technology to communicate with other road users employing future 6G networks. The communication module of a connected vehicle working at high frequencies---in the mmWave and sub-THz bands---can offer a true Gb/s experience and deliver information more efficiently than through sensor detection and processing \cite{liao2021cooperative}.

\begin{figure}[!t]
    \centering
     \includegraphics[width=0.7\columnwidth]{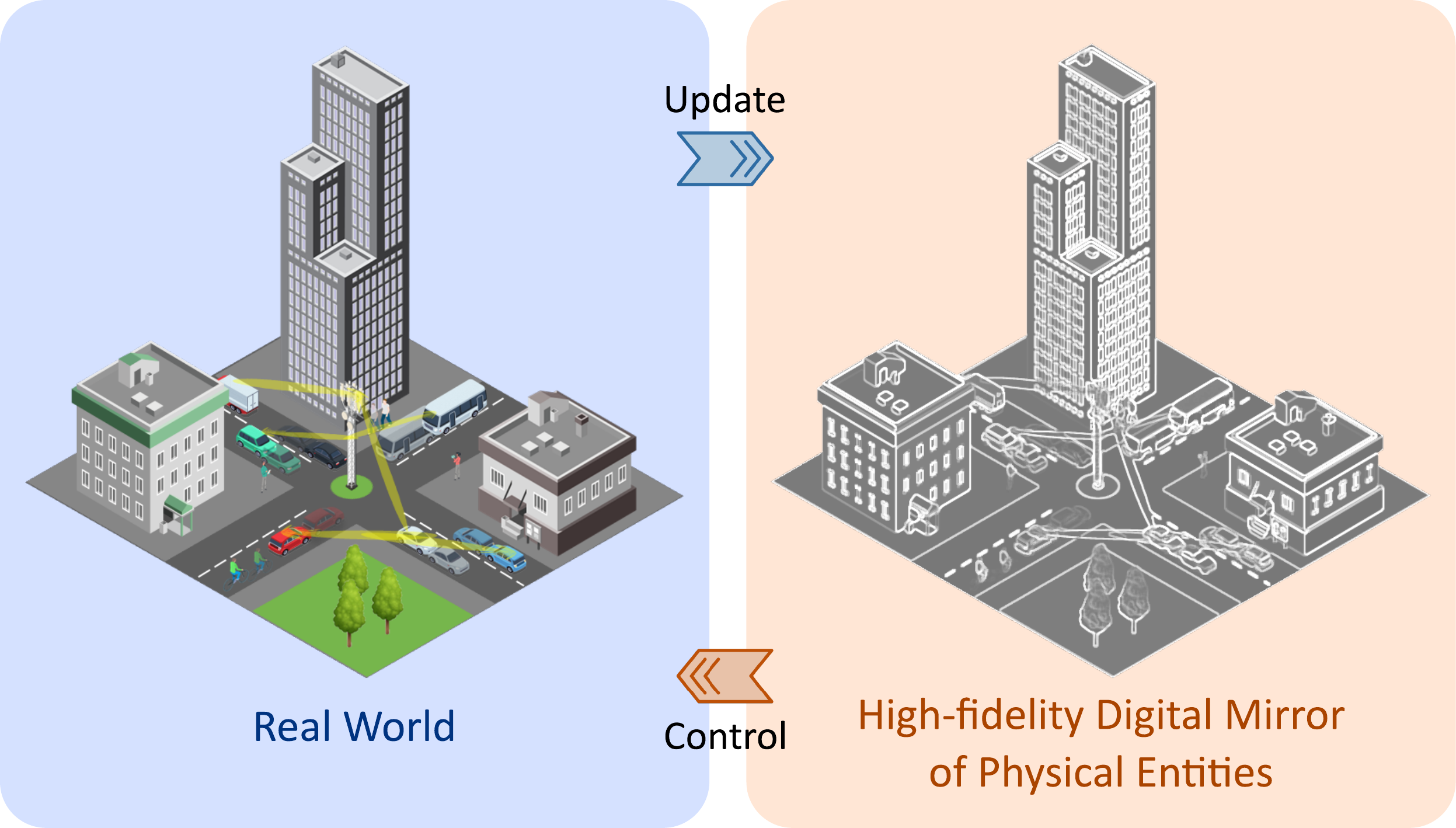}
    \caption{Interaction between a real-world V2X network and its high-fidelity digitized representation. The urban environment and the vehicles' movements govern wireless propagation in the real world. The digital replica uses high-definition reconstructed 3D maps, realistic sensor simulations, and accurate 3D ray tracing to control the communication equipment.}
    \label{fig:scenario}
\end{figure}

However, the use of higher frequency bands makes the wireless links prone to line-of-sight (LOS) blockages. The vehicles' movement and the rising density of urban buildings, which mostly determine the signal degradation, are the primary causes of discontinuous connectivity in V2X environments \cite{LinsalataLoSmap}.
In this context, quickly and opportunistically restoring and handing a direct link after a prolonged blockage is a fundamental aspect that needs to be addressed.
As a possible countermeasure, the latest 6G paradigm envisions merging the physical and digital worlds \cite{DTMagazine, alkhateeb2023real}, and it is widely acknowledged that the Digital Twin (DT) will be a key enabling technology to achieve this goal \cite{ding2022digital}.

The DT aims at providing a high-fidelity digital representation of physical phenomena. This is obtained not only by integrating simulations and available data but also by accounting for the entire phenomenon life-cycle, which provides up-to-date insights about the physical entity. Thus, a DT-enabled V2X network can get meaningful knowledge of the surrounding environment by collecting and processing a huge amount of data, and providing real-time accurate channel estimation \cite{ding2022digital}. To create such an accurate real-time digital reproduction of the physical environment, the envisioned DT has to use high-definition 3D maps and combine multi-modal sensory data from scattered devices and infrastructure nodes. The DT needs to be continuously improved to enhance its decision accuracy and to refine its approximation of the physical world, as depicted in Fig. \ref{fig:scenario}. Indeed, the demands for data reliability are stricter in dynamic scenarios than in static ones. Particularly, in scenarios with stringent time requirements and severe tolerance constraints, as for V2X networks where even a small model inaccuracy can have a remarkable impact on any decision process.

Real-time DTs have been researched in a variety of fields, including real-time remote monitoring and control in industry, risk assessment in transportation, and smart scheduling in smart cities. The use of DT in the network has mostly concentrated on operation issues for wireless communications, including edge computing, network optimization, and service management \cite{DTMagazine}.

The availability of multi-modal data for communications-related decisions (e.g., beam prediction and localization) can greatly reduce the required decision time where an exhaustive search over all the candidates would be required by the standard \cite{gu2022multimodality}. To foster the application of multi-modal datasets to decision-making in communications, Gu \textit{et al.} present in \cite{gu2022multimodality} an experimentally obtained dataset aiding beam selection in V2X mmWave bands. By contrast, our paper focuses on simulating the real vehicular world using a real-time DT, with attention to the environment's physical modelling and wireless signal propagation.

\begin{figure}[!t]
    \centering
    \begin{subfigure}[b]{0.55\columnwidth}
        \includegraphics[width=\columnwidth]{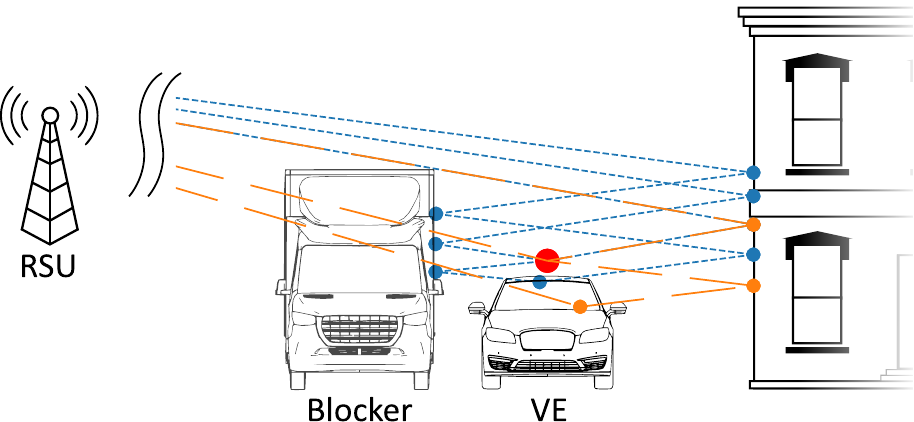}
        \caption{}
        \label{fig:blockage_example_geometry}
    \end{subfigure}\hspace{0.2cm}
    \begin{subfigure}[b]{0.3\columnwidth}
        \includegraphics[width=\columnwidth]{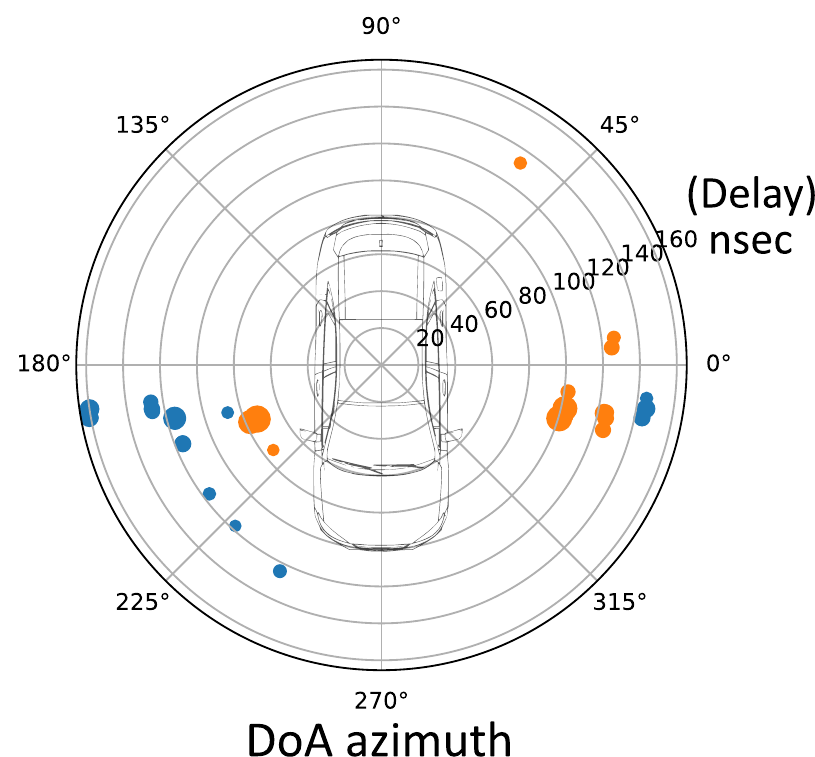}
        \caption{}
        \label{fig:blockage_example_DoA_delay}
    \end{subfigure}

    \caption[]{Geometric variation of the most powerful ray-tracing channel paths between an RSU and a VE operating at 28 GHz in an urban environment when affected (blue) or not affected (orange) by vehicular blockage from a truck in terms of (a) geometrical configuration of the main paths in the environment, and (b) directions of arrival and delays of the simulated channel paths.}
    \label{fig:blockage_example}
\end{figure}

The main contributions of this paper can be summarized as it follows.
\begin{itemize}
    \item We envision a workflow for the creation of the DT of a vehicular urban setting from multi-sensor data acquired at the communication infrastructure or ego-vehicles.
    
    \item We describe a multi-modal simulation framework for V2X wireless communications leveraging the CARLA realistic automotive 3D simulation and testing software for the generation of realistic sensor data, and the Remcom Wireless InSite accurate ray-tracing channel simulator.

    \item We showcase the dynamic ray-tracing simulation capabilities of the proposed simulation framework investigating DT-aided V2X link restoration after a prolonged communication blockage.
\end{itemize}

The rest of the paper is organized as follow. Section \ref{sect:problem_statement} defines the concept of DT-aided communications and its applicability to V2X. The proposed multi-modal simulation framework for V2X communications is highlighted in Sec. \ref{sect:method}. The experimental results and the numerical simulations for blockage handover are reported in Sec. \ref{sect:results}. Section \ref{sect:conclusion} concludes the work.

\section{Digital Twin-aided Communications}\label{sect:problem_statement}

The V2X umbrella includes the two main subcategories of Vehicle-to-Vehicle (V2V) and Vehicle-to-Infrastructure (V2I) communications. V2V, or sidelink, refers to direct communication between vehicles, without any support from the network infrastructure, while V2I includes communication between vehicles and road infrastructure, e.g., road side units (RSUs). 
Narrow-beam and high frequencies communications are foreseen as candidate solutions to high spectral and energy efficiency KPIs of future 6G V2X networks \cite{gu2022multimodality}.

We consider a multi-user V2X communication system in which the communication infrastructure and the vehicular user equipments (VEs) are capable of multi-modal sensing, i.e., they are equipped with multiple sensors (e.g., camera, LiDAR, radar) to perceive their surroundings. VEs can be also equipped with sensors to estimate their state---e.g., positioning systems and inertial measurement units (IMUs),---in terms of acting forces, position, and orientation.

Owing to highly dynamic environments, V2X communications experience rapidly varying communication link conditions, which can greatly affect communication performance and call for novel channel estimation techniques. As an example, Fig. \ref{fig:blockage_example} shows the \textit{deep} change in electromagnetic wireless propagation between a road-side unit and a VE when affected by the vehicular blockage phenomenon. The differences in terms of angles and delays of the propagation paths clearly show the variation in the spatial and temporal propagation channel patterns in the blocked and non-blocked conditions. 
As a result, dynamic environment conditions affect wireless propagation in vehicular settings, severely changing the propagation structure between the transmitter and the receiver for both V2I and V2V communications. The latter scenario is the most challenging, since both the communication nodes are dynamic in this case.

Recent contributions in the literature, e.g., \cite{jiang2022lidar, xu2022computer}, have emphasized the advantages of integrated communication and multi-modal sensing to provide effective solutions over a wide variety of tasks, including beam prediction, channel estimation, blockage prediction/hand-off, VE positioning, and object detection.
We believe that the availability of multiple sensors at the communication infrastructure and/or at VEs, in challenging channel estimation conditions, can benefit from the automotive simulation and testing framework. The latter are commonly used for the validation of Autonomous Vehicles (AV) and Advanced Driver Assistance Systems (ADAS) and enable the construction of the DT of vehicular scenarios aimed at controlling physical communication equipment based on virtual sensors simulation and accurate channel generation.

\section{A multi-modal simulation framework\\for V2X communications}\label{sect:method}

\begin{figure*}[!t]
    \centering
     \includegraphics[width=0.9\textwidth]{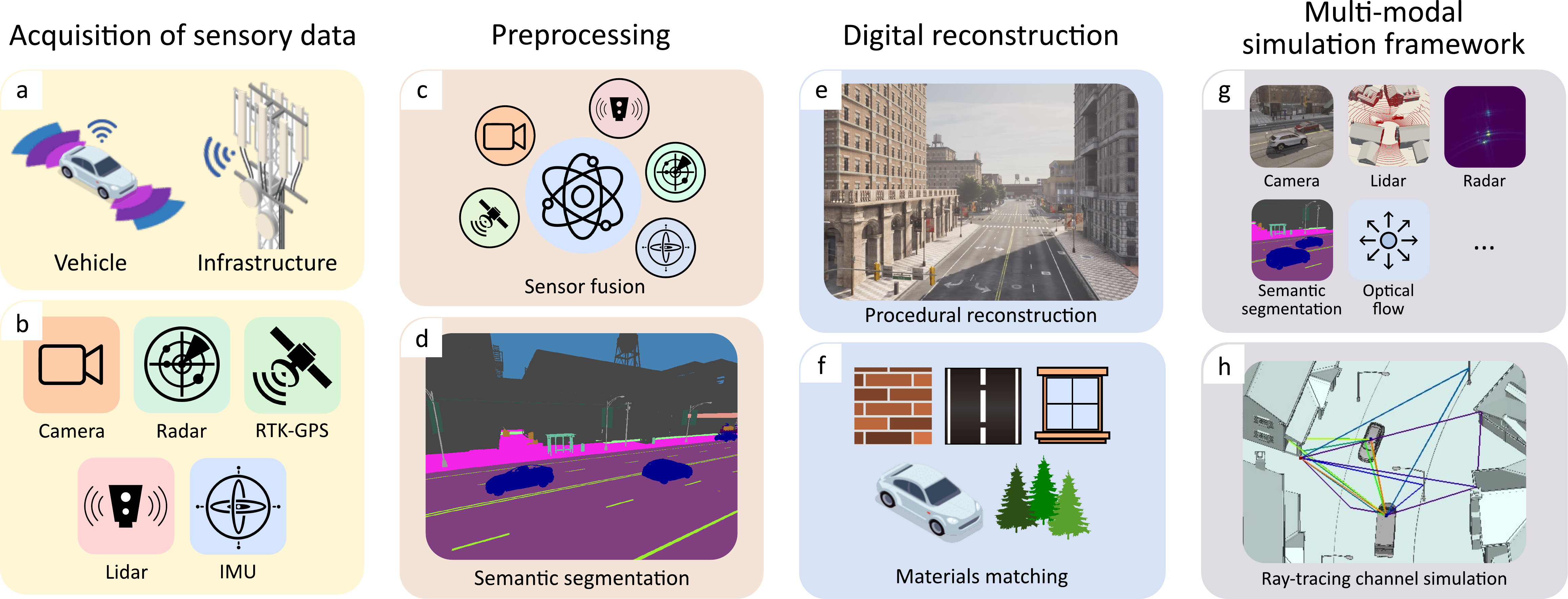}
    \caption{DT construction workflow. Acquisitions from multiple sensors at the vehicle or the communication infrastructure are preprocessed by sensor fusion and semantic instance segmentation to determine the components of the 3D scene and their materials. Procedural digital reconstruction is performed and utilized for multi-modal sensor simulation within an Unreal Engine-based automotive framework and for accurate wireless propagation simulation through a ray-tracing simulator.}
    \label{fig:workflow}
\end{figure*}

In this section, we propose a workflow for the construction of the DT of a physical urban/suburban environment and a multi-modal simulation framework to enable wireless communications. Fig. \ref{fig:workflow} depicts the proposed workflow.
The procedure is intended as an iterative approach, which constantly updates the DT from the acquired data.

\subsection{Digital Twin construction}

In the following, we briefly describe the envisioned digital twin construction workflow, introducing the sensory data acquisition, preprocessing and digital reconstruction steps, and focusing on multi-sensor simulation and ray-tracing channel generation in the reconstructed 3D environment, which are the target of this contribution.

\textbf{Sensory data acquisition} (Figs. \ref{fig:workflow}a and \ref{fig:workflow}b): the environment surrounding a sensing-capable network component (a VE or at the communication infrastructure) is perceived employing multiple sensors, e.g., camera, LiDAR, and radar. Positioning and inertial measurement systems are also common on automated vehicles, providing information on the forces acting on the vehicle, its position, and its orientation.

\textbf{Preprocessing of the acquired data} (Figs. \ref{fig:workflow}c and \ref{fig:workflow}d): the acquired data is preprocessed through sensor fusion techniques to leverage the joint availability of multiple sensor streams. Semantic instance segmentation is then performed to determine the 3D scene components---e.g., buildings, vehicles, fences, lamp posts, road marks, and vegetation. An effective method for LiDAR 3D point-cloud semantic segmentation is proposed by Zhuang \textit{et al.} in \cite{zhuang2021perception}, where the authors exploit perceptual information from RGB images and spatial-depth information from point clouds to devise a collaborative fusion scheme.

\textbf{Digital 3D reconstruction}  (Fig. \ref{fig:workflow}e): the identified 3D scene components are mapped to known 3D object models by means of procedural 3D reconstruction. Procedural modelling for urban environments leverages the patterns of architectural designs to synthesize realistic 3D models. Shape grammars \cite{stiny1971shape} have been proposed to effectively reconstruct real-environment models. In shape grammars, prior knowledge of the architectural design principles is described utilizing formal grammars. Shape grammars have been successfully integrated with data-driven processes to effectively reconstruct city models from sensor data. Nishida \textit{et al.} propose in \cite{nishida2018procedural} a system to generate a procedural model of a building from a single camera image, automatically estimating camera parameters and generating the building windows and doors geometry. Smelik \textit{et al.} provide in \cite{smelik2014survey} a more general survey on accurate procedural modelling of virtual twins, focusing on the degree of interactivity and control offered by the described methods.

\textbf{Materials matching} (Fig. \ref{fig:workflow}f): a material is associated to each 3D component class to enable the reconstructed 3D scenario for electromagnetic (EM) propagation simulations. We decompose a 3D vehicle models into bodies, windows, bumpers, wheel rims, and tires. Shape components for the 3D models within the reconstructed scenario can be derived from procedural modelling grammar. For a thorough description of the EM properties of several materials, we refer the reader to the \mbox{ITU-R} recommendation P.1238 \cite{itu_recommendation_materials}.

\textbf{Multi-modal simulation} (Fig. \ref{fig:workflow}g): the reconstructed 3D environment is integrated within the CARLA \cite{carla2017} automotive software framework, designed to foster research on Autonomous Vehicles (AV) and Advanced Driver Assistance Systems (ADAS). In particular, CARLA provides state-of-the-art realistic urban/suburban 3D simulation over different weather conditions relying on the Unreal Engine real-time game engine \cite{unrealengine}.
CARLA allows the simulation or integration of a wide variety of sensors---both static (for simulations at the communications infrastructure) and dynamic (moving jointly with the simulated vehicles):
\begin{itemize}
    \setlength\itemsep{0em}
    \item \textit{RGB camera}, provides standard coloured images and is defined for both a pinhole camera model and for real-life ultra wide-angle lenses (fisheye camera);
    \item \textit{identification/segmentation cameras}, identify objects perceived in the camera image, classifying each pixel based on the object type (semantic segmentation), separating pixels belonging to different observed objects (instance segmentation), or providing the bounding boxes associated to each object (multi-object identification);
    \item \textit{optical flow camera}, allows representing the object movements associating to each image pixel its vertical and horizontal velocity and encoding them into RGB channels;
    \item \textit{LiDAR}, provides a 3D point cloud for a given number of channels (stacked vertical lasers) and a specified horizontal angular resolution;
    \item \textit{inertial measurement unit}, provides information on the dynamic state of the vehicle to which it is attached;
    \item \textit{mmWave radar}, simulates radar sensing through specialized ray-tracing software, e.g., Remcom WaveFarer.
\end{itemize}
A multi-modal synthetic dataset for semantic segmentation based on CARLA is proposed in \cite{testolina2023selma}, where the authors show the effectiveness of deep learning model training with the proposed dataset and generalization over real-world acquired data.

\textbf{Accurate wireless channel simulation} (Fig. \ref{fig:workflow}h): to achieve an accurate simulation of the wireless propagation channel, we utilized the Remcom Wireless InSite specialized ray-tracing simulation software. Alternatively, Sionna RT, a real-time ray-tracing channel simulation software leveraging GPU hardware acceleration, has also been recently introduced. 
Ray-tracing propagation models have been proven to effectively simulate multi-path propagation and space-time channel dispersion.
The considerable progress in computing architectures and the formulation of improved computational methods enable real-time performance for multi-sensor simulations and ray-tracing wireless propagation simulation.

The reconstructed 3D model of the urban/suburban environment is imported within Wireless InSite, associating the corresponding material to each environment component.
The position and orientation of moving vehicles are modified frame-by-frame, jointly with the automotive multi-sensor simulation. Wireless InSite provides GPU acceleration and multi-threading and allows 3D ray tracing considering different types of interaction with the environment---i.e., reflections, diffraction, scattering, and transmission.
For each propagation ray between transmitter and receiver, the information provided by Wireless InSite includes the punctual details on received power, phase, direction of departure (DoD), direction of arrival (DoA), delay and Doppler shift (for moving Tx or Rx equipment).

\subsection{Datasets generation}

Besides controlling physical equipment, the proposed simulation framework enables the generation of realistic datasets of synchronized multi-sensor data. Accurate communication channels at the infrastructure and at the ego-vehicles are the basis for realistic benchmarks definition and machine/deep learning algorithms training. 

A vehicular traffic simulator---e.g, the Simulation of Urban MObility (SUMO) open-source software---can be utilized to generate realistic vehicular traffic data.
The use of a vehicular traffic simulator requires the availability of an accurate map of the road network. To this aim, we propose to extend the preprocessing phase proposed in Fig. \ref{fig:workflow} with a further step for road mark identification and HD lanes reconstruction. A multi-stage pipeline relying on clothoidal spline models is proposed in \cite{cudrano2022clothoid} for the definition of lane-level high-definition maps. We endowed the SUMO vehicle motion model with a kinematic bicycle model for the estimation of the front wheels' rotation.

Leveraging the Python-based CARLA co-simulation procedure for SUMO, we integrated the SUMO simulator with both the CARLA automotive simulation framework and with Remcom Wireless InSite to achieve dynamic multi-sensor and ray-tracing channel simulations.
The generation of vehicular traffic and wireless channel data can be integrated with network-layer simulations through event-based network simulators, like OMNeT++ and NS-3. 5G New Radio-enabled vehicular networks can be simulated, e.g., using the 5G-LENA NS-3 module or through the Simu5G OMNeT++-based simulator.

We notice that procedural 3D generation allows building a wide variety of combinations of 3D and environmental features (e.g., building 3D structures, textures, weather conditions) to generate several diversified datasets. Machine learning methods for decision-making in vehicular networks can benefit from the availability of massive diversified datasets that faithfully represent the environment dynamics to solve common vehicular communications tasks---e.g., initial access and beam prediction---at different layers in the communication stack \cite{ye2018machine}. The multitude of procedural 3D scenarios can be utilized to validate communication protocols on heterogeneous vehicular networks and diverse configurations of the wireless propagation environment.

\section{Experiments and numerical simulations}\label{sect:results}

In this section, we first demonstrate the proposed multi-modal simulation framework over an urban vehicular scenario. Then, by leveraging the ray-tracing channel DT information, we showcase the benefits of dynamic ray-tracing channel simulation in a specific and common V2X challenge: fast and opportune direct link restoration after prolonged blockage.

\subsection{Digital Twin simulations}

In this section, we demonstrate the proposed multi-modal simulation framework over an urban vehicular scenario. We assume that the sensory data acquisition, preprocessing and digital reconstruction procedures have been performed, and the 3D models of vehicles and simulation scenario are available.

\begin{figure}[!t]
    \centering
    \includegraphics[width=\columnwidth]{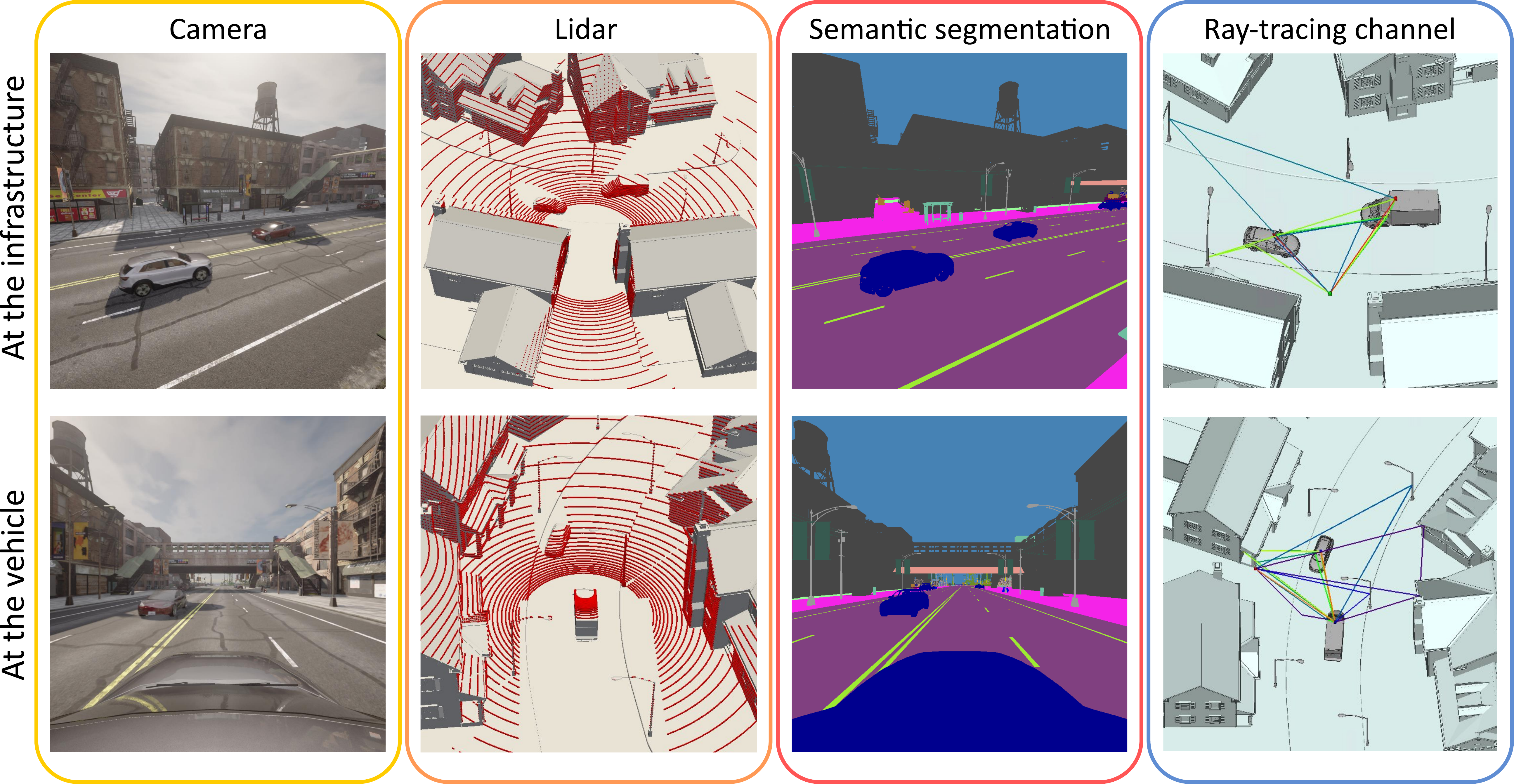}
    \caption{Simulation of RGB camera, LiDAR and semantic segmentation camera sensors and ray-tracing channel at a simulation frame for the selected urban scenario both at the communication infrastructure and at an ego-vehicle.}
    \label{fig:results}
\end{figure}

The first three columns of Fig. \ref{fig:results} present the simulation results for three different virtual sensors---i.e., RGB camera (pinhole camera model), LiDAR unit and semantic segmentation camera---in CARLA from the point of view of an RSU and of a vehicle moving in the scenario in the neighbourhood of the RSU. The camera resolution has been set up to $1920\times1080$ pixels, while the focal length is 16 mm and the camera sensor size is 36 mm x 20.25 mm. The LiDAR unit emulates a realistic radar with 64 channels, a rotation frequency of 10 Hz and a horizontal resolution of 0.6 deg, providing a 3D point cloud of the RSU and vehicle surroundings. The semantic segmentation camera annotates each RGB camera pixel with a different color for each object type, distinguishing vehicles, buildings, curbs, pavements, fences and vegetation. The figure shows the real-time realism and accuracy reached by the CARLA Unreal Engine-based automotive framework during simulation.

The generation of the ray-tracing channel through Remcom Wireless InSite is presented in the last column of Fig. \ref{fig:results} for both a transmitter at the infrastructure and at an ego-vehicle. The simulated RSU and VEs operate at 28 GHz carrier frequency. The figure shows the most powerful generated rays. Ray-tracing has been performed considering isotropic Tx and Rx antennas, and a maximum of 7 reflections, 6 diffractions and 1 scattering interactions between transmitter and receiver. The simulation of MIMO channels over antenna arrays can be achieved by applying the array response vectors or precoding/combining schemes to the simulated propagation paths according to the communication system architecture.

During the materials matching phase, concrete has been selected for buildings structures, glass for vehicles and building windows, wood for fences, perfect electric conductor (PEC) for lamp posts, vehicle bodies, vehicle bumpers and wheel rims, and rubber for vehicle tires.

\subsection{Dynamic vehicular blockage handover}

In this section, we showcase the proposed multi-modal simulation framework by addressing the problem of V2I link recovery in an urban scenario. To this aim, let us consider a set of VEs moving within the BS coverage area that perform a beam handover (BH) procedure whenever the link quality degrades owing to the a prolonged blockage condition.
At each time instant of the procedure, the optimal beamformers searched within codebooks $\mathcal{F}$ and $\mathcal{W}$ at both BS and VE sides to be used are the maximizers of the received power at the VE.

By relying on the DT information, we can find \textit{instantaneous} subsets of $\mathcal{F}$ and $\mathcal{W}$ to speed up the BH procedure, namely
\begin{equation}    \mathcal{F}_{\mathrm{sub}}\subseteq\mathcal{F},\,\,\,\,\mathcal{W}_{\mathrm{sub}}\subseteq\mathcal{W}.
\end{equation}

\begin{table}[t!]
\centering
\caption{Simulation parameters.}
\begin{tabular}{l c}
\toprule
\textbf{Parameter} & \textbf{Value(s)}\\
\noalign{\smallskip}
\hline
\noalign{\smallskip}
\footnotesize
Comm. carrier freq. &  $28$ GHz\\
Modulation & 64 QAM \\
OFDM numerology      & 4\\
Beamforming Codebook      & DFT \\
N. of VE antenna $N_\mathrm{VE}$ & 4 $\times$  2 \\\noalign{\smallskip}
Single pair test time $T$ & $62.5$ $\mu$s \\
Tx power $\sigma_s^2$ & $-10$ dBm\\
Rx noise power $\sigma_n^2$ & $-102$ dBm\\
SNR threshold $\gamma_\mathrm{thr}$ & 10 dB \\

\bottomrule
\end{tabular}
\label{tab:SimParam}
\end{table}

This BH approach can be directly compared with the gradient search used in beam tracking mode in the 5G NR standard~\cite{TS_38213}. 
The gradient search is initialized with an exhaustive search during the initial access to find the global optimum.
The strategy behind this approach is to focus the search around the azimuth and elevation angles, $\hat{\vartheta}$ and $\hat{\varphi}$, determined at the previous successful beam training instant before the blockage condition. The subset of BS angular codebooks is computed as
\begin{align} \label{eq:codebookGS}
    \Theta^\mathrm{sub}_\mathrm{BS} &= \left\{\hat{\vartheta} + \vartheta_k \bigg\lvert \,\,k = -\frac{K}{2}, \dots \frac{K}{2}\right\},\\
    \Phi^\mathrm{sub}_\mathrm{BS} &= \left\{\hat{\varphi} + \varphi_{k} \bigg\lvert \,\,k = -\frac{K}{2}, \dots \frac{K}{2}\right\},
\end{align}
where $\hat{\vartheta} +\vartheta_k\in\Theta_\mathrm{BS}$, $\hat{\varphi} +\varphi_k\in\Phi_\mathrm{BS}$ and $K^2$ represents the methods' codebook cardinality. High values of $K$ lead to higher complexity, while low values can lead to losing the global optimum and thus to error propagation. 

\begin{figure}[!t]
    \centering
    \includegraphics[width=0.50\textwidth]{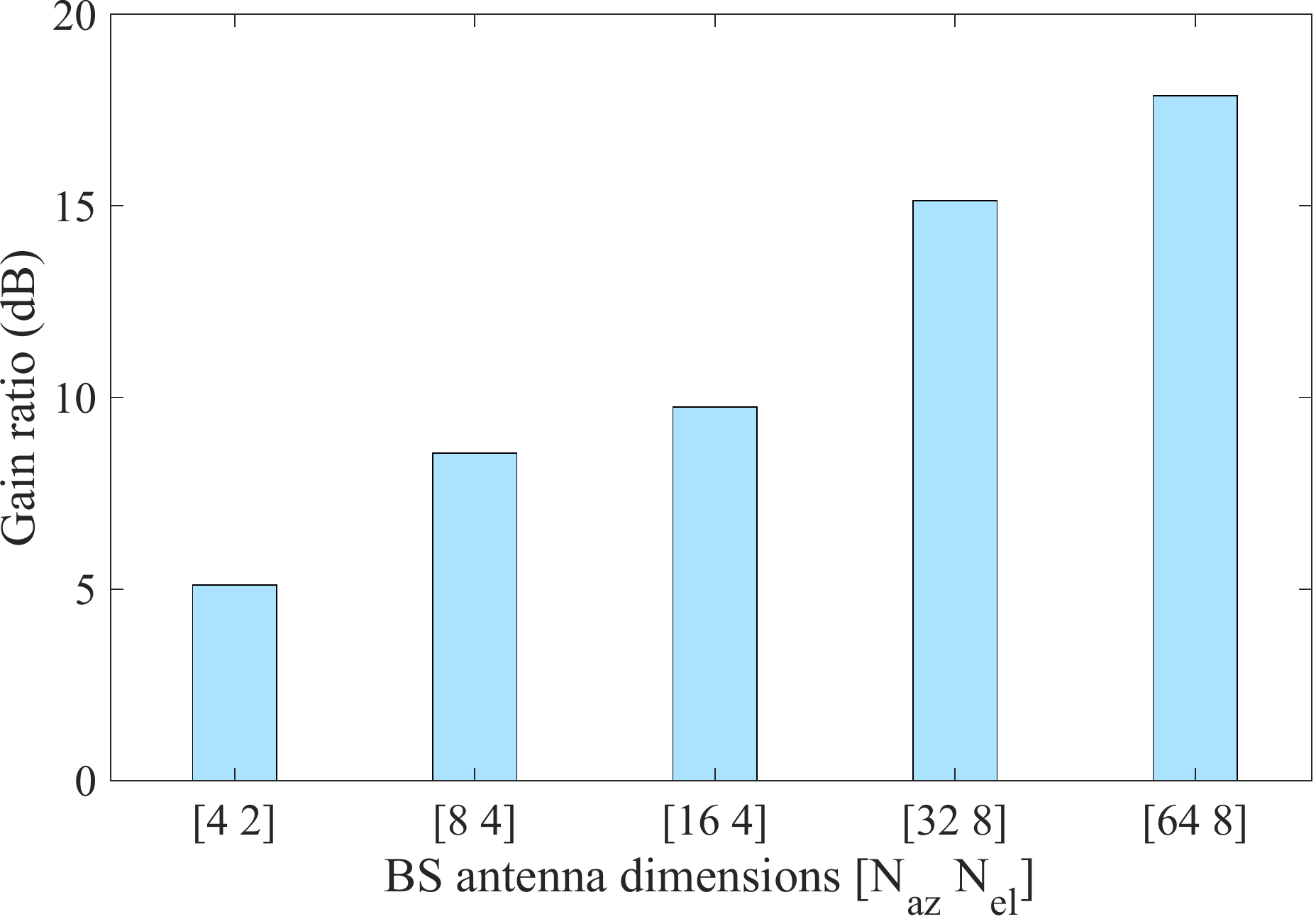}
    \caption{Beamforming gain ratio versus BS antenna array dimensions between DT-aided blockage handover and the 5G NR approach for codebook cardinality $K = 2$.}
    \label{fig:results_N}
\end{figure}

\begin{figure}[!t]
    \centering
    \includegraphics[width=0.50\textwidth]{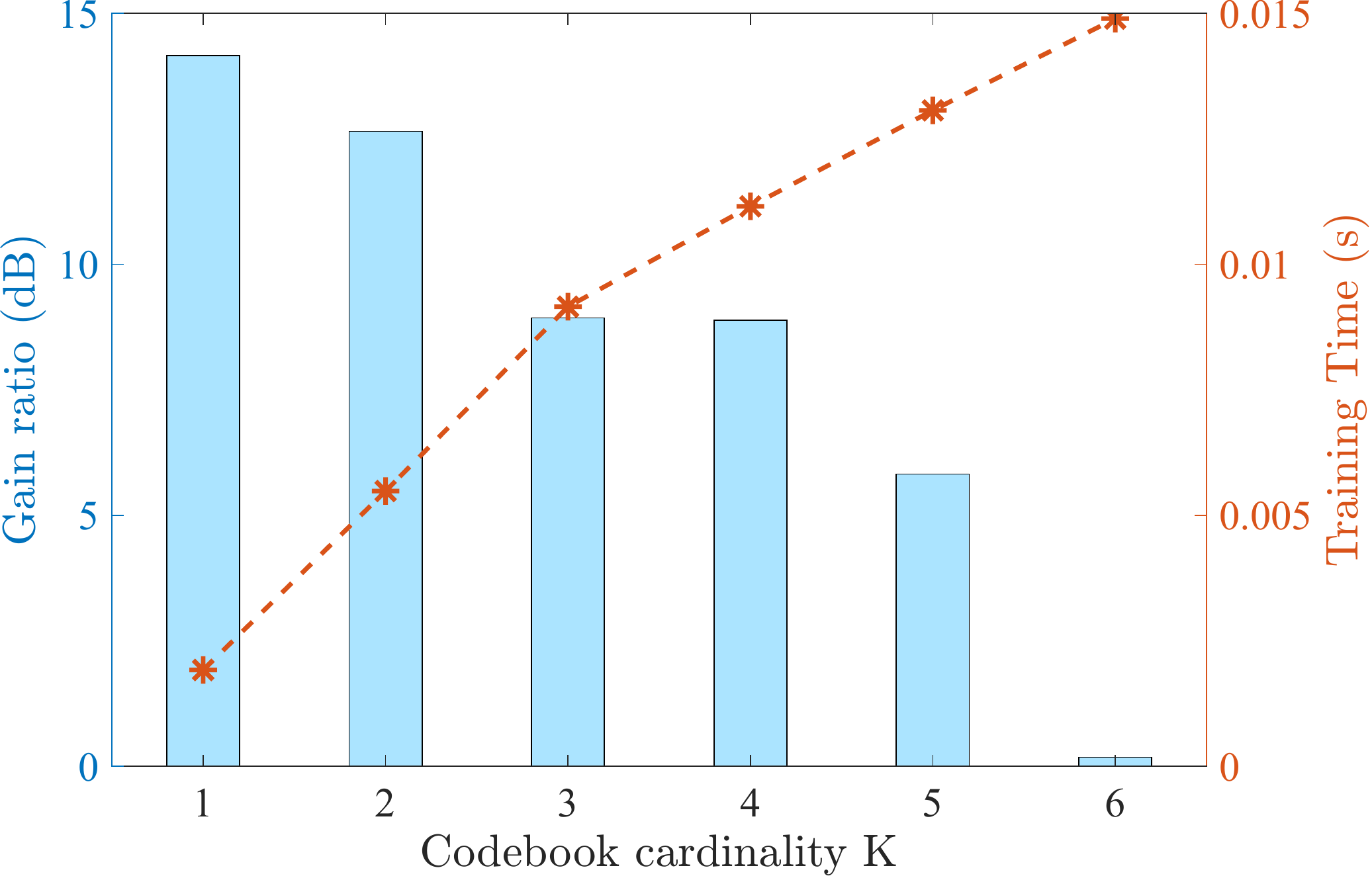}
    \caption{Beamforming gain ratio between DT-aided blockage handover and the 5G NR approach and Training Time variation for the 5G NR approach versus codebook cardinality $K$ for BS antenna array with dimensions $N_{az} = 16$ and $N_{el} = 4$.}
    \label{fig:results_K}
\end{figure}

We compare the two BH approaches through two performance metrics: the training time and the beamforming gain ratio.
The \textit{cost of performing} a search across codebooks $\mathcal{F}$ and $\mathcal{W}$ can be expressed in terms of beam training time $T_{\mathrm{train}}$, defined as
\begin{equation} \label{eq:TrTime}
    T_{\mathrm{train}} = T \, \lvert\mathcal{F}\rvert \lvert\mathcal{W}\rvert,
\end{equation}
where $T$ is the time taken for testing a single BS-VE beam pair and $\lvert\cdot\rvert$ is the cardinality of the given set. 
The \textit{beamforming gain ratio} is defined by 
the mismatch in the chosen beamformers by the two approaches. An higher gain loss means an higher degradation in the link SNR.

The main simulation parameters are reported in Table \ref{tab:SimParam}.
Fig. \ref{fig:results_N} presents the numerical results of the beamforming gain ratio between the DT-aided BH and the 5G NR approach when $K=2$, varying the number of antennas at the BS. As can be observed, the DT-aided BH achieves superior performance compared to the standard BH procedure, particularly when the codebook cardinality is high. This highlights the utility of up-to-date information provided by the twin in DT-aided BH, while the 5G NR approach requires much longer beam training to achieve comparable results.

Indeed, in Fig. \ref{fig:results_K}, we provide numerical results for the beamforming gain ratio between the DT-aided BH and the 5G NR approach and beamforming training time for the 5G NR approach under varying codebook cardinality $K$ setting the BS antenna array dimensions to $N_{az} = 16$ and $N_{el} = 4$. We remark how the difference between the two approaches diminishes as $K$ increases, nevertheless, resulting in an unavoidable increase in training time and network overhead.

\section{Conclusion}\label{sect:conclusion}
In this paper, we envisioned a novel workflow and proposed a multi-modal simulation framework for the construction of a realistic DT of a physical scenario from multi-sensor acquisitions in dynamic vehicular environments. The developed DT is aimed at improving existing V2X communications by allowing to control physical equipment based on accurate virtual sensors and channel data. Moreover, the proposed multi-modal simulation framework allows the generation of realistic multi-sensor data based on CARLA, and ray-tracing channel datasets based on Remcom Wireless InSite at the communication infrastructure and at an ego-vehicle, enabling the definition of realistic benchmarks and the training of machine/deep learning algorithms. We showcased the proposed simulation framework investigating DT-aided V2X link restoration after a prolonged communication blockage. The results show consistent behavior in the comparison of DT-aided blockage handover versus the traditional 5G NR approach, under dynamic ray-tracing channel generation providing realistic vehicular blockage simulation conditions.

\footnotesize
\section*{Acknowledgment} This work was partially supported by the European Union under the Italian National Recovery and Resilience Plan (NRRP) of NextGenerationEU, partnership on “Telecommunications of the Future” (PE00000001 - program “RESTART”, Structural Project 6GWINET) and by the Joint Lab between Huawei Milan Research Center and Politecnico di Milano. This paper was also supported by “Sustainable Mobility Center (Centro Nazionale per la Mobilità Sostenibile – CNMS)” project funded by the European Union NextGenerationEU program within the PNRR, Mission 4 Component 2 Investment 1.4.

\bibliographystyle{IEEEtran}
\bibliography{bibliography.bib}

\end{document}